\documentclass{article}
\usepackage{spconf,amsmath,graphicx}
\usepackage{subfig}
\usepackage{afterpage}

\usepackage{cite}
\usepackage{amsmath,amssymb,amsfonts, cuted,capt-of}
\usepackage{algorithmic}
\usepackage{graphicx}
\usepackage{tikz}
\usepackage{hyperref}
\usepackage{lipsum}
\usepackage{caption}
\usepackage{siunitx}
\usepackage{xcolor}

\usepackage{subfig}

\usepackage{textcomp}
\usepackage{xcolor}
\def\BibTeX{{\rm B\kern-.05em{\sc i\kern-.025em b}\kern-.08em
    T\kern-.1667em\lower.7ex\hbox{E}\kern-.125emX}}

\begin{document}

\title{Spatio-Temporal Perception-Distortion Trade-off in Learned Video SR}
\name{Nasrin Rahimi and A. Murat Tekalp \thanks{This work is supported in part by TUBITAK 2247-A Award No.~120C156 and KUIS AI Center funded by Turkish Is Bank. A. M. Tekalp also acknowledges support from Turkish Academy of Sciences (TUBA).}}
\address{ Department of Electrical \& Electronics Engineering and KUIS AI Center \\ 
Koç University, 34450 Istanbul, Turkey}
\maketitle
%\copyrightnotice

\vspace{-20pt}
\begin{abstract}
Perception-distortion trade-off is well-understood for single-image super-resolution. However, its extension to video~super-resolution~(VSR) is not straightforward, since popular perceptual measures only evaluate naturalness of spatial textures and do not take naturalness of flow (temporal coherence) into account. To~this~effect, we propose a new measure of spatio-temporal perceptual video quality emphasizing naturalness of optical flow via the~perceptual straightness hypothesis~(PSH) for meaningful spatio-temporal perception-distortion trade-off. We also propose a new architecture for perceptual VSR~(PSVR) to explicitly enforce naturalness of flow to achieve realistic spatio-temporal perception-distortion trade-off according to the~proposed measures. Experimental results with PVSR support the hypothesis that a meaningful perception-distortion tradeoff for {\it video} should account for the naturalness of motion in addition to naturalness of texture. 
%Experimental results validate the effectiveness of proposed measures and the PVSR architecture.
\end{abstract}
%---------------------------------------------------------------------------
%---------------------------------------------------------------------------
\vspace{-5pt}
\section{Introduction}
\label{sec:intro}
\vspace{-5pt}
Early works on learned video super-resolution (VSR) employed supervised training to minimize the l2/l1 loss~\cite{kap2016,cab2017}. However, it is well-known that models that are trained to minimize the mean-squared-error (MSE) result in blurry unnatural looking textures because the minimum MSE estimate is a probability weighted average of all feasible solutions. Later, generative perceptual VSR methods that optimize a weighted combination of l2/l1 loss, a no-reference adversarial loss, and full-reference perceptually motivated losses have been proposed. Perceptual VSR methods provide sharper texture in each frame of video at the expense of a decrease in PSNR as predicted by perception-distortion trade-off theory~\cite{PDT2018}. 

Typical VSR methods, whether based on fidelity only or perceptual criteria, calculate losses per frame, and therefore, do not take temporal coherence explicitly into account. Since humans can detect unnatural motion and jitter easily, temporal inconsistencies result in a video with low perceptual quality, even if the texture in each frame looks natural. Few works explicitly model or enforce the temporal consistency of frames in VSR. However, enforcing naturalness of motion remains as a non-trivial problem that should be considered within the context of spatio-temporal perception-distortion trade-off using a new well-defined measure of naturalness of motion.

To this effect, we propose a new measure to evaluate spatio-temporal naturalness of super-resolved videos for better spatio-temporal perception-distortion trade-off and present a new perceptual VSR (PSVR) model with two discriminators, where one evaluates naturalness of texture and the other naturalness of motion. 
We discuss related works in Section~\ref{sec:rw}. Section~\ref{sec:st_pdt} proposes a new measure for spatio-temporal naturalness of video based on the perceptual straightness hypothesis~\cite{henaff2019perceptual}. The proposed temporally coherent PVSR model is introduced in Section~\ref{sec:pvsr}. Evaluation methodology and comparative experimental results are presented in Section~\ref{sec:eval}. Finally, Section~\ref{sec:conclusion} concludes the paper.
\vspace{-15pt}
%---------------------------------------------------------------------------
\section{Related work}
\label{sec:rw}
\vspace{-6pt}
\subsection{Perception-Distortion Trade-off for Images}
\vspace{-5pt}
According to the perception-distortion trade-off theory~\cite{PDT2018}, distortion refers to dissimilarity between an original image~$X$ and its reconstruction $\hat X$, which is measured using a full-reference (FR) measure,
%$\Delta (X,\hat X)$, 
while perceptual quality is the degree to which $\hat X$ appears as a valid natural image, measured by a no-reference~(NR) measure, regardless of how similar it is to $X$. It was shown that an algorithm cannot be both very high fidelity and perceptually natural regardless of the distortion measure used~\cite{PDT2018}. However, extension of this result to video (temporal dimension) is nontrivial as discussed in Section~\ref{sec:st_pdt}.
\vspace{-8pt}

\subsection{Perceptual VSR Methods and Their Evaluation}
\vspace{-5pt}
Perceptual VSR models \cite{lucas2019generative,perez2018perceptual,xie2018tempogan,chu2020learning} aim to achieve a trade-off between distortion and perceptual quality using conditional GANs \cite{qiao2019image} and perceptual losses. \cite{lucas2019generative} employs VSRResNet as generator along with a discriminator, but do not address temporal consistency either~in training or in evaluation. \cite{perez2018perceptual} proposed a recurrent architecture and a video discriminator to reinforce temporal consistency. Addressing the SR problem for fluid flow, \cite{xie2018tempogan} employs a temporal discriminator in addition to a spatial discriminator. Besides a recurrent architecture, \cite{chu2020learning} introduced a spatio-temporal discriminator, called TecoGAN, together with a set of training objectives for realistic and temporally coherent VSR. Motivated by the perceptual straightening hypothesis (PSH)~\cite{henaff2019perceptual} for human vision, \cite{kancharla2021improving} proposes a quality-aware discriminator model to enforce the straightness of trajectory of the perceptual representations of predicted video frames for the video frame prediction task.
In this paper, we apply PSH to the VSR task; furthermore, we do not impose perceptual straightness as a constraint, but use it as a measure for perceptual evaluation.

VSR models are typically evaluated by averages (over frames) of PSNR and SSIM as distortion measures, and of LPIPS (FR measure)~\cite{zhang2018unreasonable} and NIQE (NR measure)~\cite{mittal2012making,6272356} as perceptual quality measures~\cite{tu2020comparative}. However, these measures are designed for single images; hence, they do not directly evaluate motion artifacts or temporal coherence of frames. 
%VMAF~\cite{VMAF} fuses multiple measures using SVM-based regression to include information fidelity, detail loss, and, mean co-located pixel difference.
As measures of temporal coherency, \cite{chu2020learning} introduces tOF, which is pixel-wise difference of estimated flow vectors, and tLP, which is the difference between LPIPS of successive frames of predicted and pristine videos. Alternatively, MOVIE~\cite{seshadrinathan2009motion} integrates both spatial and temporal aspects of distortion assessment based on a spatio-spectrally localized multiscale framework. STEM \cite{kancharla2021completely} combines the~NIQE metric with a blind temporal algorithm which is based on perceptual straightening hypothesis \cite{henaff2019perceptual}. Nonetheless, there is no study on the effectiveness and role of these measures in evaluating spatio-temporal perception-distortion trade-off in VSR.

%---------------------------------------------------------------------------
\setlength{\belowdisplayskip}{0pt} \setlength{\belowdisplayshortskip}{0pt}
\setlength{\abovedisplayskip}{3pt} \setlength{\abovedisplayshortskip}{0pt}
\vspace{-8pt}
\section{New Measures for Spatio-Temporal Perception-Distortion Trade-off in VSR} 
\label{sec:st_pdt}
\vspace{-7pt}
%The perception-distortion trade-off also exists in VSR; however, 
The role of naturalness of motion in perceptual evaluation of VSR models has not been well-studied. To fill this gap, we propose new spatio-temporal perception and distortion measures to evaluate the perception-distortion trade-off in VSR. 
\vspace{-22pt}
\subsection{\textbf{Spatio-Temporal Perceptual Measure}}
\vspace{-5pt}
The spatio-temporal perceptual quality can be evaluated as a combination of spatial and temporal naturalness measures. \vspace{2pt}\\
%Subjective evaluation of perceptual quality is costly. Hence, 
{\bf Spatial Naturalness:} For still frames, it is typical to use LPIPS~\cite{zhang2018unreasonable} 
%which measures the perceptual similarity between two images using deep features 
or NIQE~\cite{mittal2012making,6272356} measures, which estimate the deviation of image statistics from that of natural images. For video, spatial perceptual quality $PQ_{Spatial}$ can be evaluated by averaging LPIPS or NIQE over all frames. For example,
\begin{equation}
     PQ_{Spatial}=\frac{1}{N} \sum_{n=1}^{N}LPIPS_{Alex}({X}_{n} , \hat{X}_{n})
\end{equation}
\vspace{0pt}

\noindent {\bf Temporal Naturalness:} We are inspired by the perceptual straightening hypothesis~\cite{henaff2019perceptual}, which states that the human visual system transforms visual stimuli to a perceptual domain, where natural sequences follow a straighter temporal trajectory. A two-stage computational model that imitates the nonlinear properties of the early visual system, namely, the retina, lateral geniculate nucleus (LGN) and V1, to transform sequences into perceptual domain was proposed in~\cite{henaff2019perceptual}. Let's consider an $N$-frame video $\{X^{n}\}_{n=1,...,N}$. The~perceptual representations $\{P^{n}\}_{n=1,...,N}$ are high-dimensional vectors in the perceptual space. The curvature $Cur(n)$ at node (frame) $n$ is defined as the angle between successive displacement vectors, $V^{n} =  P^{n} - P^{n-1}$, $n=2,3,...,N$, which is computed by the dot product of vectors $V^{n}$ and~$V^{n+1}$
\begin{equation}
    Cur(n)=arccos\left(\frac{ V^{n} \cdot V^{n+1}}{\parallel V^{n}\parallel\parallel V^{n+1}\parallel}\right) \vspace{3pt}
\end{equation}
The straightness of the trajectory at frame $n$ in the perceptual space is given by $ST(n) = \pi - Cur(n)$. The straightness of the trajectory in the intensity space can be computed similarly. 

The average difference between the straightness of a sequence in the intensity domain $ST^{I}$ and the perceptual domain $ST^{P}$ is defined as a measure of temporal naturalness
%\vspace{-4pt}
\begin{equation}
PQ_{Temporal}=\frac{1}{N} \sum_{n=2}^{N}(ST^{p}(n)-ST^{I}(n))  \vspace{3pt}
\end{equation}
\noindent {\bf Spatio-temporal Naturalness:}
A natural sequence will have higher $PQ_{Temporal}$ and lower $PQ_{Spatial}$. Consequently, a spatio-temporal perceptual quality measure can be defined as 
\begin{equation}
\label{p_st}
     P_{ST}=\frac{ PQ_{Spatial}}{ PQ_{Temporal}} \vspace{3pt}
\end{equation} 
where lower scores indicate better quality.
\vspace{-10pt}

\subsection{\textbf{Spatio-Temporal Distortion Measure}}
\vspace{-4pt}
The classical FR measure of fidelity is pixel-wise MSE, which is applied to video by averaging over all $N$-frames, as
\begin{equation}
  MSE_{Pix} = \frac{1}{N} \sum_{n=1}^{N}\|X_{n}-\hat{X}_{n}\|_{2}^{2} \vspace{3pt}
\end{equation}
where $X_{n}$ denotes frame $n$. To take temporal distortions into account explicitly, we also define optical flow MSE by
\begin{equation}
\label{O_MSE}
     MSE_{OF}=\frac{1}{N}\sum_{n=2}^{N}\|OF(X_{n},X_{n-1})-OF(\hat{X}_{n},\hat{X}_{n-1})\|_{2}^{2}
\end{equation}
where $OF(X_{n},X_{n-1})$ is the flow between frames $X_{n}$ and $X_{n-1}$. This is similar to $t_{OF}$ measure defined in terms of $l1$ distance~\cite{chu2020learning}.
We define a spatio-temporal distortion measure $D_{ST}$ for video as a weighted sum of $MSE_{Pix}$ and $MSE_{OF}$ 
\begin{equation}
\label{Distortion}
     D_{ST}= MSE_{Pix}+ \alpha MSE_{OF} \vspace{3pt}
\end{equation}
The parameter $\alpha$ is chosen empirically considering that the effect of optical flow distortion should neither be neglected nor dominate the overall distortion measure. We set $\alpha = 1000$.
\vspace{-5pt}

%---------------------------------------------------------------------------
\section{A New Perceptual VSR Architecture \\ For Natural Texture and Motion}
\label{sec:pvsr}
\vspace{-5pt}
%VSR methods upscale a given low-resolution video $Y_{t}$ by a factor of $s$, such that the super-resolved video $\hat X_{t}$ resembles $X_{t}$ by some metrics. 
%\cite{PDT2018} shows that GANs provide a principled way to approach the perception-distortion bound for image restoration.
We propose a new GAN-based perceptual VSR (PVSR) architecture motivated by the spatial and temporal naturalness measures discussed in Section~\ref{sec:st_pdt}.
The proposed PVSR architecture, depicted in Figure~\ref{fig:model3}, employs a generator model, a~flow estimation model, and two discriminators. 
Generator models that minimize a distortion-only loss yield superior fidelity measures, such as PSNR, at the expense of lower perceptual scores. Inclusion of an adversarial texture loss encourages perceptually more realistic textures; however, this causes motion artifacts due to inconsistent hallucinations in successive frames. We show that an additional adversarial motion loss encourages perceptually more natural videos.
\vspace{-10pt}

\subsection{PVSR Generator Model}
\vspace{-5pt}
The PVSR model can use any generator network~$G$. In this paper, we experimented with two VSR models: EDVR \cite{wang2019edvr}, which processes each picture independent of the previous outputs, hence, is more prone to motion artifacts, and BasicVSR++~\cite{chan2022basicvsr++}, which is a recurrent model, hence, inherently generates temporally more coherent frames. The~resulting two PVSR models are called PEDVR and PBasicVSR++, respectively. Table 2 shows that we can get significant improvements in the perceptual scores even with PBasicVSR++.
\vspace{-20pt}

\subsection{PVSR Texture and Motion Discriminators}
\vspace{-5pt}
To achieve both natural texture and motion, the PVSR model employs two discriminators: a spatial~(texture) discriminator $D_{S}$ to distinguish ground-truth (GT) frames from reconstructed ones, and a temporal~(motion) discriminator~$D_{T}$ to discriminate between optical flows estimated from the~GT frames and those estimated from reconstructed SR frames. 
%\vspace{-10pt}
\begin{figure}[t!]
	\centering
	\includegraphics[width=0.95\columnwidth]{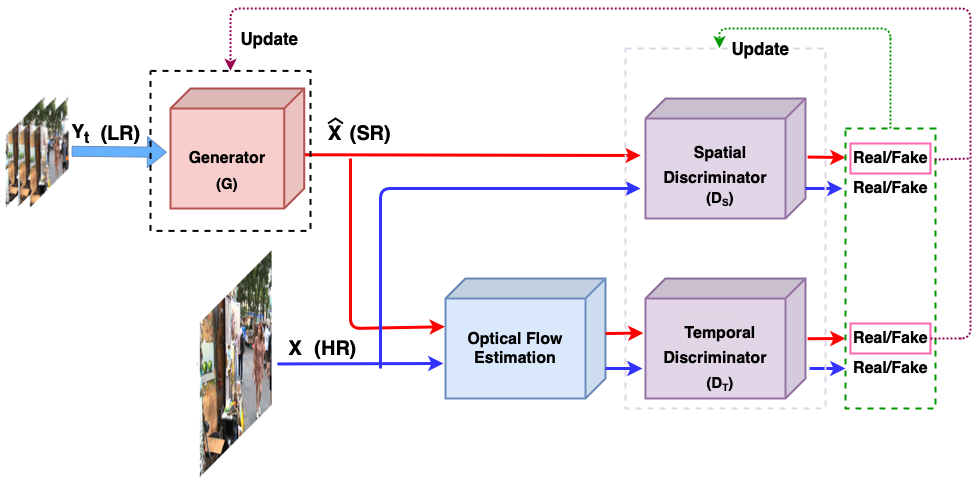} \vspace{-5pt}
	\caption{\small The proposed perceptual VSR (PVSR) model architecture. PEDVR and PBasicVSR++ are two instances of PVSR architecture.}   \vspace{-10pt}
	\label{fig:model3}
\end{figure}
The spatial and temporal discriminators have the same architecture, containing five convolutional layers. The first layer has 3$\times$3 kernel and stride 1, while the others have 4$\times$4 kernels and stride 2. The number of filters in each layer is 64, 64, 64, 128, and 256 respectively. The~final feature map is input to a fully connected layer to compute the fake/real scores. The input to the flow discriminator is the optical flow between the current and previous frames estimated by the pre-trained PWC-Net~\cite{sun2018pwc}, depicted with a blue box in Figure~\ref{fig:model3}.
\vspace{-10pt}

\subsection{\textbf{Loss Functions}}
\vspace{-5pt}
In order to ensure naturalness of both texture and motion, we propose the generator loss function to include both texture loss $\mathcal{L}_{G,Spatial}$ and motion loss $\mathcal{L}_{G,Temporal}$: \vspace{2pt}
\begin{equation}
    \label{Total_loss2}
    \mathcal{L}_{G}=\mathcal{L}_{G,Spatial}+\mathcal{L}_{G,Temporal}
\end{equation}
\vspace{-7pt}

\noindent {\it \textbf {Naturalness of Texture:}}
Inspired by SRGAN \cite{ledig2017photo}, we use the~following texture loss \vspace{-1pt}
\begin{equation}
\label{Texture_loss}
\mathcal{L}_{G,Spatial}=\lambda_{1}\mathcal{L}_{Pix}+\lambda_{2}\mathcal{L}_{vgg}+\lambda_{3}\mathcal{L}_{Pix,adv}  \vspace{3pt}
\end{equation}
where $\mathcal{L}_{Pix}$ is the pixel-wise $l_2$ loss to ensure texture fidelity, $\mathcal{L}_{vgg}$ is $l_2$ loss of the GT and SR feature maps extracted from a pre-trained VGG-19 network, and the adversarial loss $\mathcal{L}_{Pix,adv}$ aims to maximize the probability that the spatial discriminator will be fooled by the generator given by:
\begin{equation}
\label{pix_adv}
    \mathcal{L}_{Pix,adv}=-\frac{1}{N}\sum_{n=1}^{N}\log(D_{S}(\hat{X}))  \vspace{3pt}
\end{equation} 
where $N$~denotes the number of samples in a mini-batch.
\vspace{3pt}
%---------------------------------------------

\noindent {\it \textbf{Naturalness of Motion:} }
To ensure that the reconstructed video has coherent motion, the temporal loss is defined by
\begin{equation}
    \label{flow_loss2}
    \mathcal{L}_{G,Temporal}=\lambda_{4}\mathcal{L}_{Flow} +\lambda_{5}\mathcal{L}_{Flow,adv}   \vspace{5pt}
\end{equation}
where $\mathcal{L}_{Flow}$ is the $l_2$ loss between optical flows $OF^{X}$ and  $OF^{\hat X}$ estimated from GT frames and SR frames, respectively. $\mathcal{L}_{Flow,adv}$ denotes the adversarial flow loss similar to~\ref{pix_adv}, which is calculated based on the output of flow discriminator~$D_{T}$ and encourages the generator to produce sequences with a natural flow.
Furthermore, to improve the long-term temporal consistency and avoid temporal accumulation of artifacts, we exploit Ping-Pong loss introduced in~\cite{chu2020learning}.
%----------------------------------------------------------------------------
\vspace{-8pt}

\section{Evaluation}
\label{sec:eval}
%\vspace{-6pt}
%We present the training details, comparison with the state of the art, and an ablation study using different evaluation measures and the perception-distortion trade-off.
%\vspace{-5pt}

\begin{table}[b!]
\centering
\caption{\small{Loss terms/weights for each model: Cb, CS, and BCE denote Charbonnier, cosine similarity, and binary cross entropy loss.}}
\vspace{-9pt}
\label{tab_loss}
\resizebox{\columnwidth}{!}
{
\begin{tabular}{|c|c c c c c c c c|c c|}
\hline
          &  \bf  $\mathcal{L}_{Pix}$  & $\mathcal{L}_{vgg}$ & $\mathcal{L}_{Pix,adv}$ & $\mathcal{L}_{Flow}$ & $\mathcal{L}_{Flow,adv}$ &  $\mathcal{L}_{pp}$ & $\mathcal{L}_{warp}$& $\mathcal{L}_{offset}$  &  lr  &  iter  \\ 
\hline\hline
 \bf EDVR \cite{wang2019edvr}    & Cb / 1     &  $\times$     &  $\times$      & $\times$       &  $\times$     & $\times$      &  $\times$  &$\times$ & $4\times10^{-4}$ & 600K+600K  \\  
 \bf PEDVR   & $l_2$ / 1 & $l_2$ / 1   &   BCE / 0.1  &  $l_2$ / 0.2 &   BCE / 0.1 & $l_2$ / 0.5 &  $\times$  &0.008   & $5\times10^{-5}$ & 50K
   \\ 
  \bf Ablation1   & $l_2$ / 1 & $l_2$ / 0.5 &   BCE / 0.1  & $\times$       &  $\times$     & $l_2$ / 0.5 &  $\times$  & 0.008 & $5\times10^{-5}$   & 50K   \\ 
  \bf Ablation2   & $l_2$ / 1 & $l_2$ / 0.5 &   BCE / 0.1  &  $l_2$ / 0.2 &  $\times$     & $l_2$ / 0.5 &  $\times$  & 0.008   & $5\times10^{-5}$ &   50K  \\ 
 
 \hline
  \bf BasicVSR++ \cite{chan2022basicvsr++}      & Cb / 1     &  $\times$     &  $\times$      & $\times$       &  $\times$     & $\times$      &  $\times$  &$\times$  & $1\times10^{-4}$ & 600K   \\  
  \bf PBasicVSR++  & Cb / 1 & $l_1$ / 1   &   BCE / 0.05  &  MSE / 0.1 &   BCE / 0.05 & $\times$ &  $\times$  & $\times$   &  $5\times10^{-5}$ & 200K
   \\ 
  \bf Ablation1   & Cb / 1 & $l_1$ / 1 &   BCE / 0.05  & $\times$       &  $\times$     & $\times$ &  $\times$  &  $\times$  &$5\times10^{-5}$  &  200K  \\ 
  \bf Ablation2   & Cb / 1 & $l_1$ / 0.5 &   BCE / 0.05  &  MSE / 0.1 &  $\times$     & $\times$ &  $\times$  &$\times$  &  $5\times10^{-5}$  &  200K  \\ 
 \hline
  \bf TecoGAN  & Cb / 1    & CS / 0.2    &   BCE / 0.01 & $\times$       &  $\times$     & Cb / 0.5    &  Cb / 1 & $\times$  &  $5\times10^{-5}$  & 500K   \\ 
\hline
\end{tabular}
}
\end{table}
\vspace{-10pt}
\begin{table*}[t!]
\centering
\caption{\small{Quantitative comparison of methods on REDS4. The best score is marked \textcolor{blue}{blue}, while best score within each group is in {\bf bold}.}}
\vspace{-6pt}
\label{tab_results}
\resizebox{0.8\textwidth}{!}{
\begin{tabular}{|c|c c c c c c c c c  |}
\hline
          &  PSNR(dB)/SSIM $\uparrow$  &   Pixel MSE $\downarrow$  &   OF MSE ($\times100$) $\downarrow$   & NIQE $\downarrow$ &LPIPS(Alex)$\downarrow$ &LPIPS(Vgg)$\downarrow$  & Straightness ($\times100$) $\uparrow$  & $D_{ST} \downarrow$  & $P_{ST} \downarrow$  \\ 
\hline\hline
  \bf EDVR \cite{wang2019edvr} & {\bf 30.52 / 0.870} & {\bf 65.28}    &  {\bf 1.0268}  &   4.491 & 0.1982 & 0.2225 & 29.42 & {\bf 75.55} &0.6737 \\ 
  \bf PEDVR (ours)   & 28.38 / 0.829 & 105.08   & 1.1297  &  {\bf  \textcolor{blue}{ 2.788}} & {\bf 0.1064 } & {\bf 0.1929}  & {\bf 30.33} &  116.38 & {\bf 0.3508} \\
  \bf Ablation1   & 28.27 / 0.826       & 108.97   & 1.1780    &   2.886 &0.1075  & 0.1932 &  29.72 &  120.76 &  0.3617  \\
  \bf Ablation2   & 28.29 / 0.826       & 107.99   &   1.1270  &   2.876 & 0.1079  &  0.1931  & 29.81 &  119.26 &   0.3619\\ 
 
 \hline
  \bf BasicVSR++ \cite{chan2022basicvsr++}   & {\bf \textcolor{blue}{32.39 / 0.907}} & {\bf  \textcolor{blue}{43.89}}   & {\bf \textcolor{blue}{ 1.0157}}& 3.909  & 0.1302 & 0.1726 & 30.85 & {\bf  \textcolor{blue}{54.047}} & 0.4220 \\ 
   \bf PBasicVSR++ (ours)   & 31.24 / 0.897 &  57.05   &  1.0177  &  {\bf 3.193}  &{\bf \textcolor{blue}{ 0.0845}}  & {\bf \textcolor{blue}{ 0.1452}} &{\bf  \textcolor{blue}{31.44}} & 67.227   &{\bf \textcolor{blue}{ 0.2687}}\\
  \bf Ablation1   & 31.22 / 0.886     & 57.18   &   1.0396  & 3.224   & 0.0904 & 0.1497 & 31.32 &    67.576     &  0.2886 \\
  \bf Ablation2   & 31.25 / 0.887    & 56.98     &  1.0190  &   3.232  &0.0861 & 0.1457& 31.28 &  67.170       &   0.2752 \\ 
 \hline
  \bf TecoGAN \cite{chu2020learning}  & 28.13 / 0.824 & 111.76   &   1.1350  &   2.959  & 0.0949  & 0.1802 & 29.81 &  123.11 &0.3183\\
 \hline
\end{tabular}
}
\end{table*}

%\begin{figure}[h]
%	\centering
%	\includegraphics[width=\columnwidth]{temporal_comparison.png}
%	\caption{\small{Sources of motion artifacts, results on REDS4 dataset (a) EDVR, (b) PEDVR, (c) PMEDVR , %(d) Ground Truth.}}
%	\label{fig:Spatial_Naturalness}
%\end{figure}

\subsection{Training Details}
\vspace{-6pt}
We trained all models on 
%REalistic and Diverse Scenes (REDS) 
REDS dataset \cite{NTIRE2019}, which contains videos with large and complex motions. Similar to other methods trained on REDS, four clips (000, 011, 015 and 020 - called REDS4) are utilized as the test set, and the remaining 266 clips are used for training. For training, $64\times64$ patches from the LR frames are randomly cropped and super-resolved with a factor of $4$. 
The~Adam optimizer with $\beta_{1} = 0.9$, $\beta_{2}=0.99$ is utilized for training. All experiments are performed on a single NVIDIA~Tesla~V100.
\vspace{-24pt}

\subsection{Ablation Studies on PVSR} 
\vspace{-6pt}
In order to show the benefit of using the second discriminator and different loss terms on the performance of PEDVR and PBasicVSR++, we conducted two ablations: Ablation~1 model employs only a texture discriminator, trained by the loss function~(\ref{Texture_loss}). Ablation~2 model uses the flow loss in addition to the loss function in Ablation~1, and the complete PVSR models employ both discriminators trained by the loss function~(\ref{Total_loss2}). 
The pretrained moderate~EDVR and BasicVSR++ models are used as the starting point for the generators in all experiments. The weight of different loss terms, learning rates, and number of iterations for all experiments are summarized in Table~\ref{tab_loss}. The results are discussed in Section~\ref{results}.
%\vspace{-10pt}

\subsection{Comparative Results}
\label{results}
\vspace{-5pt}
We compare the PEDVR model vs. original EDVR and PBasicVSR++ vs. original BasicVSR++. We also compare both PVSR models vs. TecoGAN (as the state-of-the-art temporally coherent perceptual VSR model). For fair comparison, we also trained TecoGAN on the REDS dataset from scratch\footnote{The models and video results can be found at 
\url{https://github.com/KUIS-AI-Tekalp-Research-Group/Perceptual-VSR}.}
%\begin{figure}[b!]
%\vspace{-5pt}
	%\centering
	%\includegraphics[width=\columnwidth]{scatter.png} \vspace{-22pt}
	%\caption{Spatio-temporal perception-distortion trade-off comparison using proposed metrics:  our model %vs. 2 benchmarks}
	%\label{fig:PDt_scatter}
%\end{figure}

First let's discuss which metrics correlate better with video quality. Clearly, results of BasicVSR++ (evaluated as still frames) look more natural than those of EDVR; yet, EDVR results have better NIQE scores than BasicVSR++. This is because the~NR measure NIQE improves as the~amount of hallucinated texture increases regardless of fidelity, and the~recurrent framework of BasicVSR++ limits unrestricted hallucination in successive frames. Hence, we decided to use LPIPS in $P_{ST}$ rather than NIQE because we believe LPIPS is better correlated with the quality of VSR frames. In terms of naturalness of motion, we observe that the Straightness measure correlates well with visual quality and OF MSE. 

Inspection of Table~\ref{tab_results} shows that both EDVR and BasicVSR++ (optimized solely on pixel-wise Charbonnier loss) achieve better PSNR and OF MSE, while NIQE and LPIPS scores are worse compared to their PVSR counterparts. Ablation~1 models in both cases (with only texture discriminator) achieve better NIQE and LPIPS at the expense of worse PSNR as expected \cite{PDT2018}.
Ablation~1 models also have worse OF MSE, since hallucinated textures in successive frames lead to  inconsistent optical flow compared to the ground-truth. Introducing $\mathcal{L}_{Flow}$ term in the loss function leads to lower OF MSE in Ablation~2 models, while introducing the second (motion) discriminator leads to the best LPIPS and straightness scores achieved by the 
proposed PVSR models. 

We also note that  PBasicVSR++ model is clearly superior to TecoGAN (state-of-the-art temporally coherent perceptual VSR model) and PEDVR in all mentioned measures due to two reasons: First, we employ two separate discriminators to allow learning spatial or temporal naturalness distributions separately as opposed to a single spatio-temporal discriminator. Second, BasicVSR++ is a stronger generator model compared to EDVR and the FRVSR network used in TecoGAN.

%exploits warping loss as well as feeding a single spatio-temporal discriminator with a sequence of frames to force the generator to learn the temporal consistency. 
%\vspace{-15pt}
\begin{figure}[h]
	\centering
	\includegraphics[width=\columnwidth]{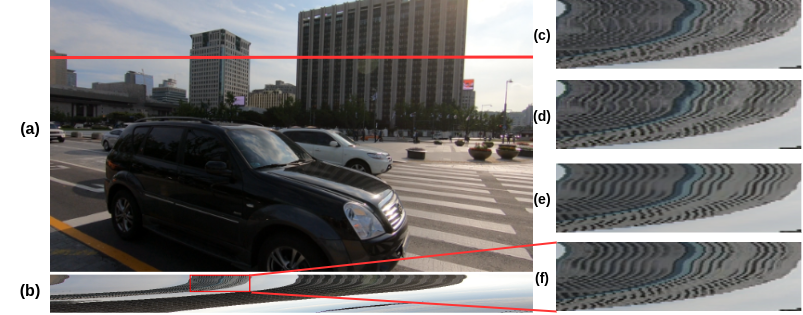}  \vspace{-15pt}
\caption{\small{(a) Video from REDS4 dataset, (b) Ground-truth temporal profile for the red line; (c)-(f) zoom in to red box for: (c) Ablation1 (PEDVR), (d)~PEDVR, (e) PBasicVSR++, (f) Ground-truth.}}
	\label{fig:profile}  \vspace{-10pt}
\end{figure}

In order to demonstrate the naturalness of motion, Figure~\ref{fig:profile} depicts temporal profiles extracted from 100 sequential frames. Comparison of zoom-ins in (c)-(f) shows the superiority of PBasicVSR++ model over all other models, 
and also that PEDVR has less temporal artifacts compared to Ablation~1 model, which does not use temporal losses.

Finally, the spatio-temporal perception-distortion trade-off can be evaluated in terms of the proposed $D_{ST}$ and $P_{ST}$ measures, which capture the spatio-temporal distortion and perceptual quality of a video. Table \ref{tab_results} shows that the original EDVR and BasicVSR++ models have the best $D_{ST}$ and worst $P_{ST}$ compared to their PVSR counterparts and their ablations. This is while, the combination of losses in \ref{Total_loss2} enables proposed PVSR models to gain the best $P_{ST}$ and strike the desired spatio-temporal perception-distortion trade-off compared to the original models as well as their ablation models.

%The proposed metrics $D_{ST}$ and $P_{ST}$ are used to evaluate different methods in terms of spatial-%temporal perception-distortion trade-off shown in Figure \ref{fig:PDt_scatter}. This plot shows that the %second discriminator forces temporal naturalness and enables our PVSR model to strike a better balance %between distortion and perception both spatially and temporally.
%On the other hand, TecoGAN exploits warping loss as well as feeding the spatio-temporal discriminator with %a sequence of frames to force it learn the temporal consistency.
%-----------------------------------------------------------------------
\vspace{-2pt}
\section{Conclusion}
\label{sec:conclusion}
\vspace{-8pt}
It is well-accepted in the community that there is no single measure of image/video quality that correlates well with human preferences. Furthermore, commonly used image perception measures such as LPIPS and NIQE, do not reflect naturalness of motion in videos. To this effect, we propose the perceptual straightness as a measure of motion naturalness and also propose a new PVSR model with two discriminators, where a flow discriminator encourages naturalness of motion. As a result, this paper advances the state of the art in spatio-temporal perception-distortion tradeoff in VSR.

Some lessons learned from this study include: 
i) a strong generator model is the most important factor to obtain the best perceptual results, ii) the perceptual quality/scores for the best model (BasicVSR++) can still be significantly improved by our PVSR architecture, 
iii) NIQE scores do not correlate well with visual VSR quality, iv) perceptual straightness measure correlates well with motion naturalness,  and v) the second (motion) discriminator improves the straightness scores.

\clearpage

\bibliographystyle{IEEEbib}
\bibliography{refs}

\end{document}